\documentclass[twocolumn,prapplied,longbibliography]{revtex4}

\usepackage[T1]{fontenc}
\usepackage{graphicx}

\usepackage{amsmath}
\usepackage{amssymb}
\usepackage{times}
\usepackage{mhchem}
\usepackage{hyperref}

\newcommand{\vavg}{\langle v_F \rangle}

\begin{document}

\title{Fermiology of two-dimensional titanium carbide and nitride MXenes}

\author{Mohammad Bagheri$^1$}
\author{Rina Ibragimova$^2$}
\author{Hannu-Pekka Komsa$^{1,2}$}
\affiliation{
$^1$Microelectronics Research Unit, University of Oulu, Oulu, Finland \\
$^2$Department of Applied Physics, Aalto University, Aalto, Finland.
}

\begin{abstract}
  MXenes are a family two-dimensional transition metal carbide and nitride
  materials, which often exhibit very good metallic conductivity
  and are thus of great interest for applications in, e.g., flexible electronics,
  electrocatalysis, and electromagnetic interference shielding.
  However, surprisingly little is known about the fermiology of MXenes, i.e,
  the shape and size of their Fermi-surfaces, and its effect on the material properties.
  One reason for this may be that MXene surfaces are almost always
  covered by a mixture of functional groups,
  and studying Fermi-surfaces of disordered systems is cumbersome.
  Here, we study fermiology of four common Ti-based
  MXenes as a function of the surface functional group composition.
  We first calculate the effective band structures of systems with explicit
  mixed surfaces and observe gradual evolution in the filling of
  the Ti-d band and resulting shift of Fermi-level.
  We then demonstrate that these band structures can be
  closely approximated by using pseudohydrogenated surfaces,
  and also compare favorably to the experimental
  angle-resolved photoemission spectroscopy results.
  By modifying the pseudohydrogen charge we then proceed to plot
  Fermi-surfaces for all systems and extract their properties, such
  as the Fermi-surface area and average Fermi-velocity.
  These are in turn used to evaluate the electrical conductivity with
  the relaxation time fitted to experimentally measured conductivities.
\end{abstract}

\date{\today}

\maketitle

\section{Introduction}

MXenes are a large class of two-dimensional (2D) materials
of transition metal (M) carbides and nitrides (X)
\cite{Naguib12_ACSNano,Gogotsi19_ACSNano}.
Distinct from other 2D materials,
MXenes are synthesized by etching out layers of atoms
from a layered bulk precursor phase. As a result of the etching process,
the dangling bonds on the surface are passivated by
a mixture of organic groups from
the etching solution, such as -O, -OH, and -F \cite{Naguib11_AM}.
MXenes are metallic and exhibit very high conductivity
in addition to good mechanical properties and easy solution
processing. As a result, they have shown great promise
for applications in conductive inks, battery electrodes,
electromagnetic interference shielding, and various types of sensors
\cite{Shahzad16_Sci,Anasori17_NRevMats,Fu19_CR,Lee20_JES}.

Given the large interest on these materials,
surprisingly little effort has been devoted to understanding
the fundamental properties of their Fermi-surfaces, i.e.,
fermiology, 
and its effect on the material properties \cite{Matsuda07_JPCM}.
Experimentally, Fermi-surfaces are commonly investigated using
angle-resolved photoemission spectroscopy (ARPES)
\cite{Matsuda07_JPCM,Lasek21_SSR}.
ARPES results for delaminated Ti$_3$C$_2$, the prototypical MXene,
were reported by Schultz et al. \cite{Schultz19_CM},
but since the measured film consists of randomly oriented flakes,
the ARPES data was azimuthally averaged, which
prevented the extraction of the Fermi-surface shape.
Quasiparticle interference within scanning tunneling microscope
(STM-QPI) is also often used to gain insight on the Fermi-surface,
but requires high quality surface with small density of scattering
centers \cite{Chen17_JPCM} --- a luxury not afforded in the case of MXenes,
where the surfaces are covered by a random mixture of organic groups
and other residues from the synthesis.
In fact, to the best of our knowledge, clean STM images of MXene
surfaces have not been reported in the literature.
Finally, Fermi-surfaces could be probed via de Haas van Alphen effect
\cite{Ouisse17_MRL},
but again no reports exist for MXenes.

Also computational studies of Fermi-surfaces are very rare,
even when the band structures have been reported in numerous papers
and are relatively well understood \cite{Xie13_PRB,Khazaei13_AFM,Khazaei17_JMCC}.
Hu et al. calculated the Fermi-surface of monolayer and
bulk Ti$_2$C$_2$(OH)$_2$ for two different layer stackings \cite{Hu15_SRep}.
In addition, some insight can be gained from the Fermi-surfaces of
the precursor MAX phases, which were presented in Ref.\
\onlinecite{Ouisse17_MRL}.
While the Fermi-surfaces for pure O or OH terminated structures
are straightforward to evaluate,
since the surfaces are passivated by a mixture of groups
(stabilized by strong nearest neighbor interactions between the groups
\cite{Ibragimova19_ACSNano,Ibragimova21_JPCL})
and the electronic structure of O and OH-terminated surfaces
differ markedly, it is not clear how the Fermi-surface
evolves with changing functional group composition.
Calculations for the mixed group surfaces are cumbersome due to
requiring the use of supercells and the resulting band folding.
The band structures can be unfolded with the effective band
structure method, where each state from the supercell Brillouin zone
is projected to each of the folded planewaves from
the primitive cell Brillouin zone \cite{Ku10_PRL,Popescu10_PRL}, 
but smearing of the bands still complicates the analysis.

In this paper, we show how to overcome these problems and
report results for the Fermi-surfaces of
Ti$_3$C$_2$, Ti$_2$C, Ti$_4$N$_3$, and Ti$_2$N
monolayers with varying O/OH composition.
We calculate the effective band structures for mixed group surfaces
and show that they can be well approximated by pseudohydrogenated surfaces.
We analyze the Fermi-surface properties as a function of
the surface group composition and compare the evaluated
electrical conductivities to the experiments.
We also compare the Ti$_3$C$_2$ band structure to those
obtained from recent ARPES experiments.

\section{Methods}

All density-functional theory calculations are carried out
using VASP software \cite{kres2,kres3},
together with projector augmented plane wave method \cite{PAW}.
We adopt the Perdew-Burke-Ernzerhof exchange-correlation functional for solids (PBEsol)\cite{PBEsol}.
The plane wave cutoff is 550 eV and
the Brillouin zone of the primitive cell is sampled with a
16$\times$16 k-point mesh.

The effective band structures (EBS) are calculated using
BandUP software \cite{Bandup1}. 
The atomic structures for the mixed surfaces were determined
via combination of cluster expansion and Monte Carlo simulations
in our previous publication \cite{Ibragimova21_JPCL}, and
we use the same 4$\times$4 supercell 
special quasi-ordered structures
as constructed therein.

Electronic band structures are interpolated using BoltzTrap2 software,
which is also used for the integration of the transport properties
\cite{Boltztrap2}.
We use k-point mesh interpolation factor 200 and temperature 300 K.
Fermi-surfaces are analyzed and plotted using the ifermi software \cite{ifermi},
with interpolation factor 20.

\section{Results}

\subsection{Band structures of mixed surface MXenes}

\begin{figure*}[!ht]
\begin{center}
  \includegraphics[width=16cm]{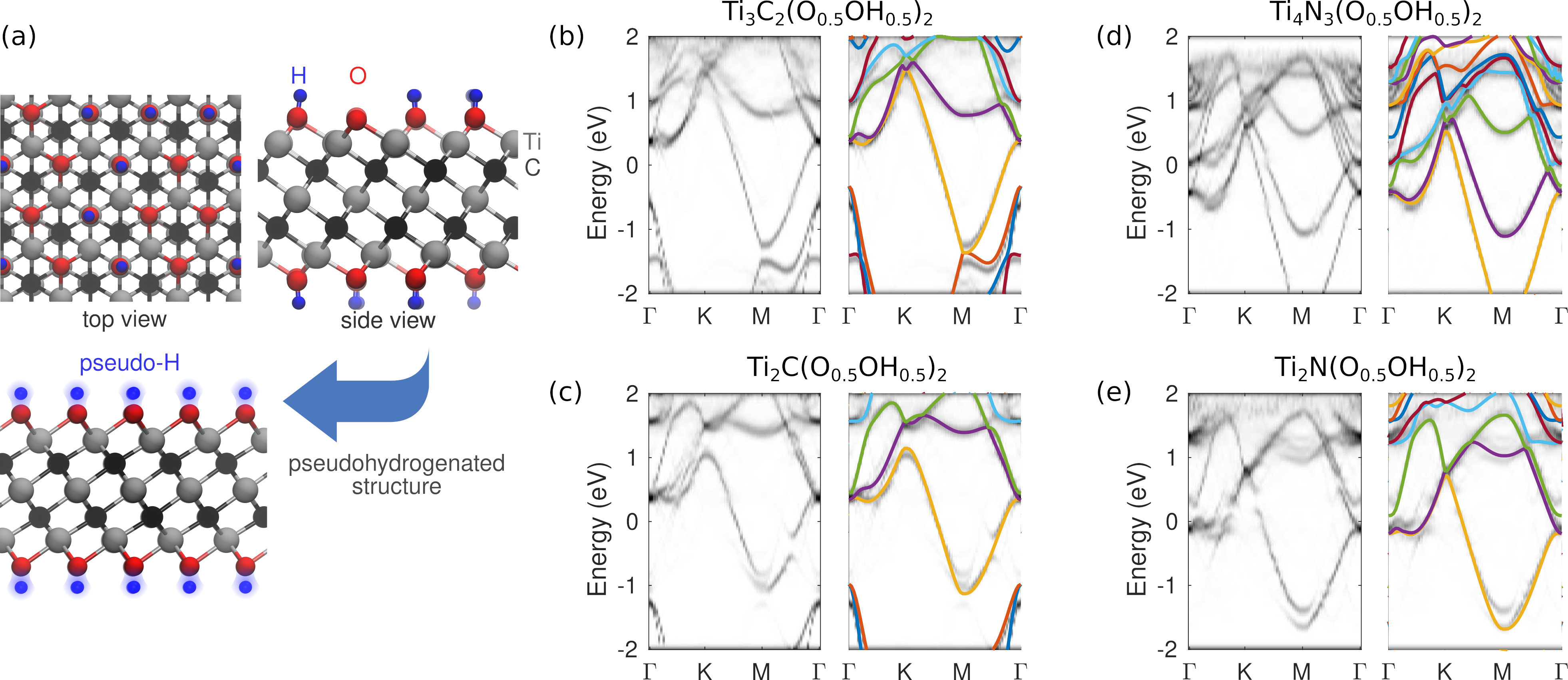}
\end{center}
\caption{\label{fig:ebs}
  (a) Atomic structure for Ti$_3$C$_2$ with a surface functionalized
  by a mixture of O and OH groups (top) and the corresponding
  pseudohydrogenated structure (bottom). 
  (b-e) Effective band structures for all materials (left panel)
  and band structures of pseudohydrogenated surfaces overlaid
  on the EBS (right panel). Fermi-level is set to zero.
}
\end{figure*}

More realistic structural models for the distribution of functional
groups were developed by us in
Refs.\ \onlinecite{Ibragimova19_ACSNano,Ibragimova21_JPCL}.
As an example, the structure of Ti$_3$C$_2$ with O$_{0.5}$OH$_{0.5}$
surface composition is shown in Fig. \ref{fig:ebs}(a).
In the left panels of Fig. \ref{fig:ebs}(b-e)
we show the effective band structures for
the four Ti-based MXenes with O$_{0.5}$OH$_{0.5}$ surface composition.
From these results, and from comparison to the band structure of
pure surfaces (see Figs.\ \ref{fig:bs1}--\ref{fig:bs2} below),
it is clear that the band structures 
undergo a gradual evolution and there are no
extraneous bands arising from the surface groups.
The band(s) at the Fermi-level arise from
the non-bonding t$_{\rm 2g}$ states of Ti-d character \cite{Hu17_JPCC}.
Consequently, to a very good approximation, the surface groups only control the
electron concentration and consequently the Fermi-level position in the metal d-bands,
as has been previously assumed \cite{Khazaei13_AFM,Khazaei17_JMCC}.
This also suggests that other charge acceptor or donor species could be
used just as well for controlling the Fermi-level position.

Given that effective band structures are unwieldy for detailed studies of
these materials' fermiology, we propose to use pseudohydrogenated surfaces
to control the Fermi-level position. 
In practice, we use the primitive cell of a fully OH-terminated structures,
but with the H atom replaced
by a pseudohydrogen atom with atomic number $Z=0.25$, $0.50$, or $0.75$
(i.e., proton charge is $Ze$ and total electron charge $-Ze$).
The pseudohydrogen band structures for $Z=0.5$ are shown on the right panels
of Fig.\ \ref{fig:ebs}(b-e), overlaid on top of the effective
band structures.
In all cases, the band structures of pseudohydrogenated MXenes
agree extremely well with the EBS,
thereby further corroborating the physical picture described above.

It is worth noting that we also calculated the band structures for
systems with one side fully terminated by O and the other by OH,
but the resulting band structures did not agree favorably
with the EBS due to broken symmetry.
Moreover, such approach would obviously be limited to
a O$_{0.5}$OH$_{0.5}$ composition.

\begin{figure*}[!ht]
\begin{center}
  \includegraphics[width=14cm]{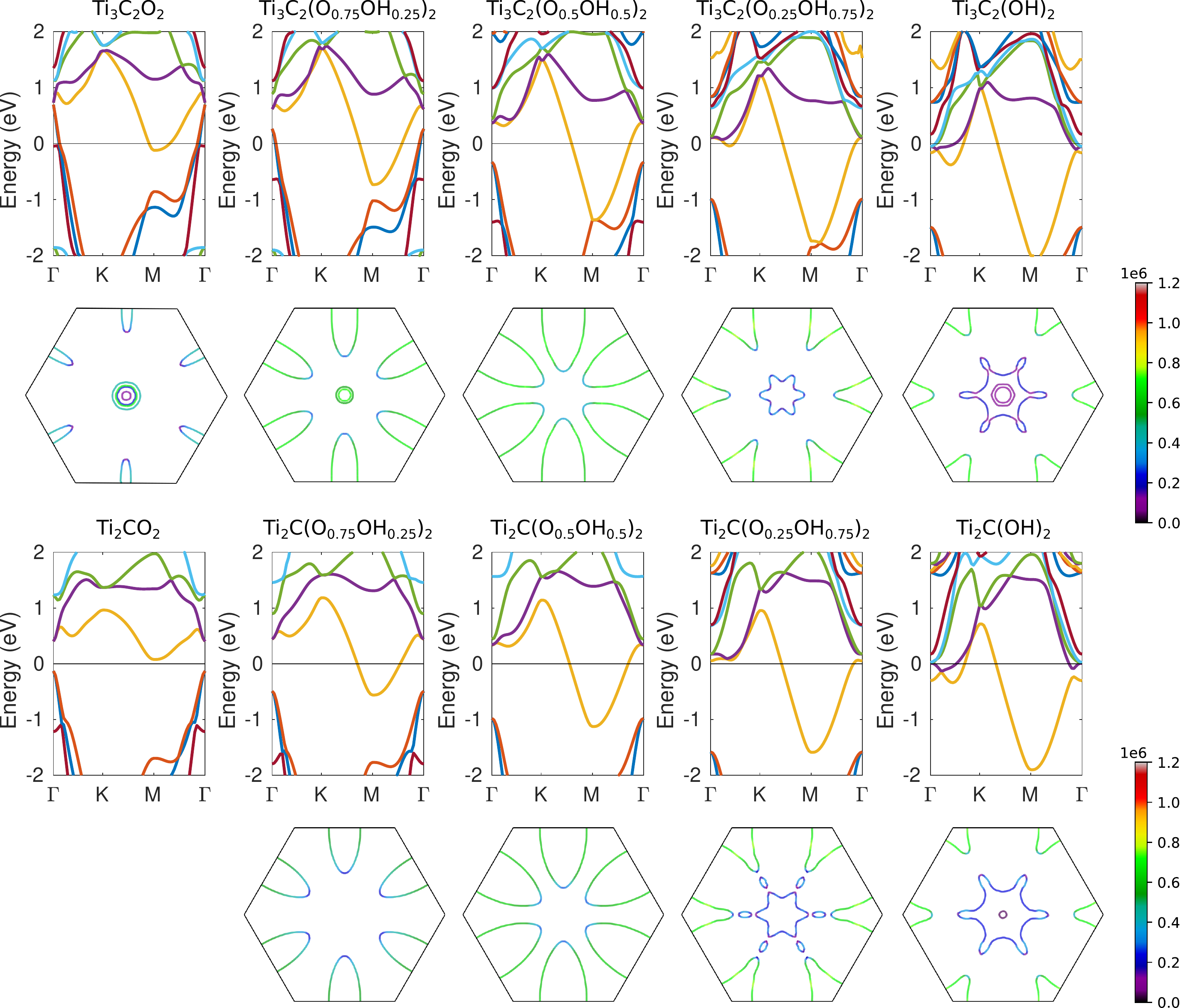}
\end{center}
\caption{\label{fig:bs1}
  Evolution of the band structures of Ti$_3$C$_2$ and Ti$_2$C with
  surface group composition.
  The Fermi-surfaces are plotted under each panel, with the color
  denoting Fermi-velocity.
}
\end{figure*}

\begin{figure*}[!ht]
\begin{center}
  \includegraphics[width=14cm]{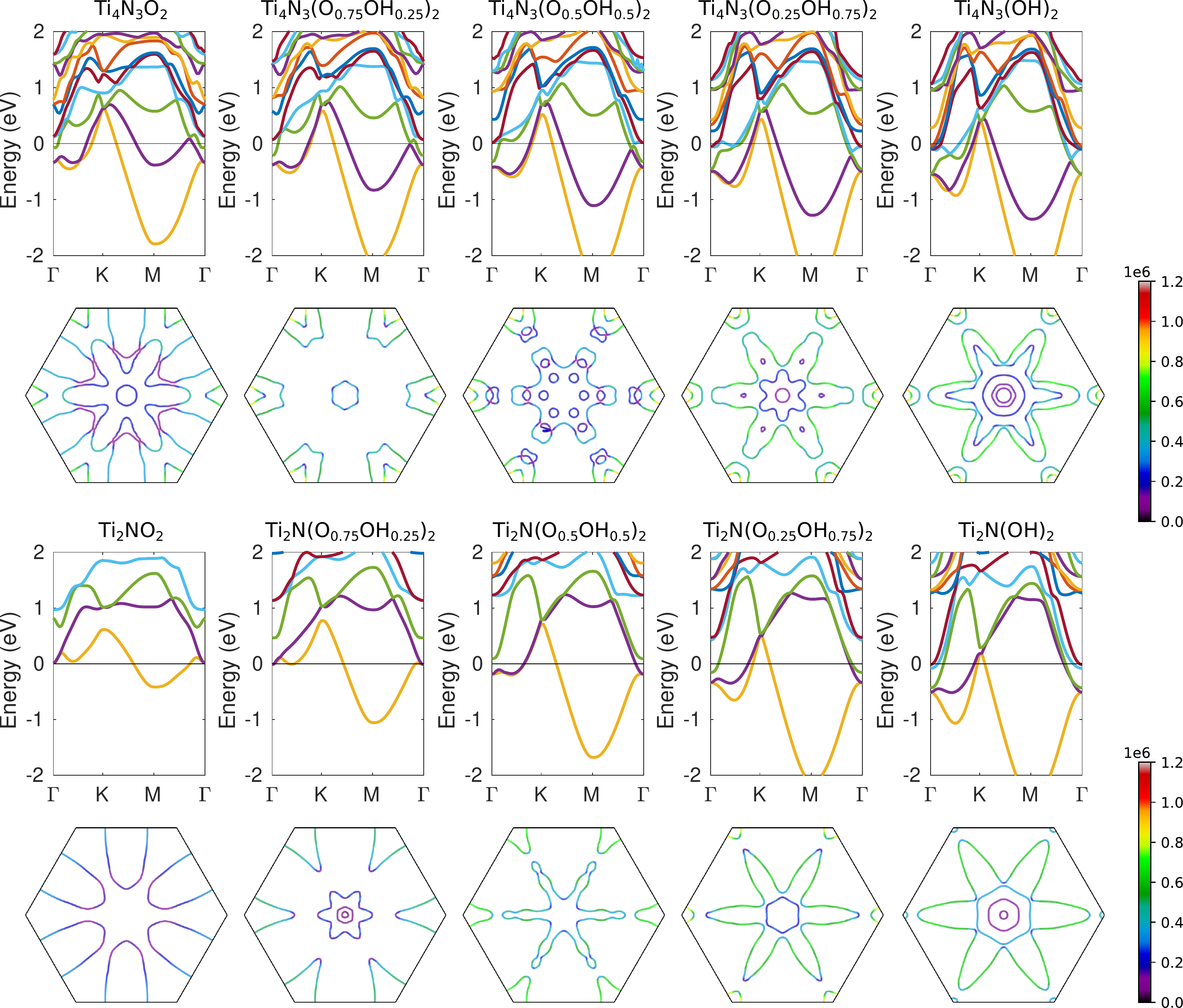}
\end{center}
\caption{\label{fig:bs2}
  Evolution of the band structures of Ti$_4$N$_3$ and Ti$_2$N with
  surface group composition.
  The Fermi-surfaces are plotted under each panel, with the color
  denoting Fermi-velocity.
}
\end{figure*}

The band structure evolution with O/OH composition,
for all materials and five different O/OH compositions,
are collected in Fig.\ \ref{fig:bs1} and \ref{fig:bs2}.
From these plots, the gradual change in the band energies
is easily resolved. 

Based on the analysis by Hu et al. \cite{Hu17_JPCC}, the states below Fermi-level
(red colored band and below in Figs.\ \ref{fig:bs1} and \ref{fig:bs2})
arise from bonding state between Ti-d and C-p/N-p ($e_{\rm g}$),
and similar to bulk carbides and nitrides \cite{Haglund93_PRB}.
The states at and above Fermi-level (orange colored band and above)
arise from non-bonding $t_{\rm 2g}$ states of Ti-d.
In most cases there is a gap between these two band manifolds,
but the Fermi-level resides in the Ti-d manifold in all cases
except for the fully O-terminated Ti$_2$C.
Due to the larger electronegativity of N, as compared to C,
the Ti-N bonding states are clearly lower in energy and lead
to a larger gap below the nonbonding states.
The Ti-C bands cross the Fermi-level around $\Gamma$-point only
for O-rich Ti$_3$C$_2$, but thereby prevent it from becoming semiconductor
as happens with Ti$_2$CO$_2$.

Due to the additional electron(s) in N, as compared to C,
there is one additional electron per X atom to fill the M-d bands
and consequently Fermi-level resides higher within this band.
For instance, the lowest energy band (orange) is exactly half-filled
by the one additional electron in the case of Ti$_2N$,
whereas the band is empty in the case of Ti$_2$C.
Similar half-filling of the band can be obtained in the case of
O$_{0.5}$OH$_{0.5}$ surface composition of Ti$_3$C$_2$ or Ti$_2$C.



The band fillings can also be readily obtained via electron counting.
In the case of Ti$_3$C$_2$O$_2$ and Ti$_2$CO$_2$,
Ti, C, and O are in the oxidation states +4, -4, and -2, respectively,
as in TiC and TiO$_2$.
Upon addition of H atoms (oxidation state +1) and the subsequent
occupation of the non-bonding Ti-d band,
the charge balance is achieved by decreasing the
oxidation state of Ti.
The nitrogen atoms (oxidation state -3) have a similar effect.

%

O/OH composition controls the occupation of the t$_{\rm 2g}$ band, but since
it is nonbonding, it should have relatively minor effect on the energetics.
Among the purely terminated surfaces, O-terminated surfaces have
been predicted to be the most stable, but the mixed surfaces arise
in practice due to the strong attractive interactions
between the opposite type surface groups \cite{Ibragimova19_ACSNano,Ibragimova21_JPCL}.
Consequently, since the surface composition is dominated by
the interactions between surface groups, it is rather independent
on the M and X species, as we observed in Ref.\ \onlinecite{Ibragimova21_JPCL}
(we also note, that the energy gain upon mixing is not captured in the
pseudohydrogen models).
%
%
Even if the O/OH mixing cannot be easily avoided,
we expect that the Fermi-level position
can still be tuned by introducing alloying in the M and X sublattices.

\subsection{Fermi-surface properties}

The Fermi-surfaces are plotted below the band structures in
Figs.\ \ref{fig:bs1} and \ref{fig:bs2}, and colored by the Fermi-velocity.
When Fermi-level crosses the lowest energy Ti-d band,
it leads to hexagonally symmetric ellipsoids around the M-points.
These have fairly large Fermi-velocities, reaching values above
$0.5\cdot 10^6$ m/s, which is only about factor of two lower than in the highly
conducting elemental metals.
At large OH concentrations (or any concentration in the case of  Ti$_4$N$_3$)
when the Fermi-level rises to cross the set
of states around $\Gamma$-point, the Fermi-surfaces develop more complex features.

\begin{figure*}[!ht]
\begin{center}
  \includegraphics[width=15cm]{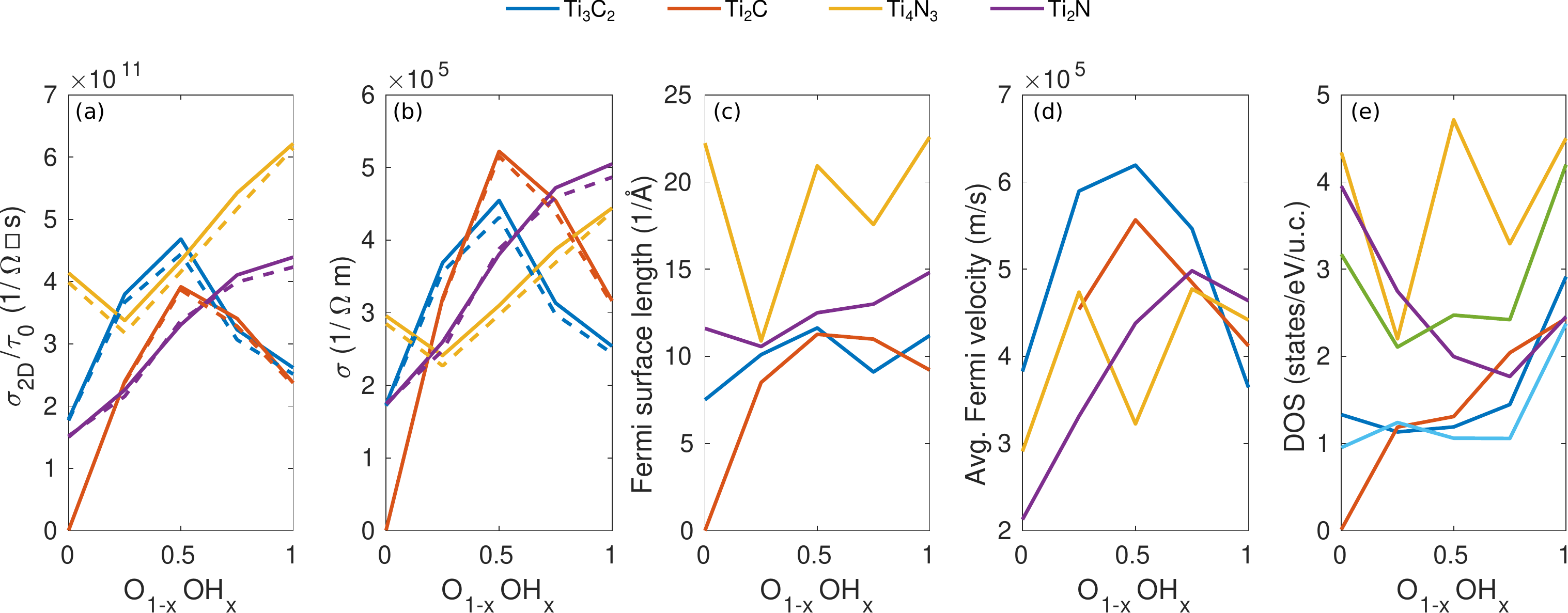}
\end{center}
\caption{\label{fig:props}
  Fermi-surface properties: (a) sheet conductivity,
  (b) conductivity, (c) Fermi-surface length,
  (d) Fermi-velocity, and (e) density of states.
}
\end{figure*}

With access to Fermi-surfaces of (effectively) mixed surface MXenes,
we can next estimate their properties.
Since MXenes are known to exhibit excellent electrical conductivity,
we evaluated the sheet conductivity within the constant relaxation time
approximation (in x-direction) as
\begin{subequations}
\begin{align}
  \sigma_{\rm 2D}
  &= \frac{e^2}{2\pi^2\hbar} \tau_0 \int_{E=E_F} \frac{v_x^2(k)}{|v(k)|} dk_F \label{eq:sigmaa}\\
  &= \frac{e^2}{2\pi^2\hbar} \tau_0 \frac{\vavg}{2} \cdot l_F \label{eq:sigmab}
\end{align}
\end{subequations}
see Supplemental Material for more detailed description.

The sheet conductivities are shown in Fig.\ \ref{fig:props}(a).
The solid lines show the results from the full integral (Eq.\ \ref{eq:sigmaa})
and the dashed lines from average Fermi-velocity approximation (Eq.\ \ref{eq:sigmab}).
All four considered MXenes show similar $\sigma_{\rm 2D}/\tau_0$, mostly falling
within 2--4 $1/\Omega\square$s.
The approximation (Eq.\ \ref{eq:sigmab}) works very well and thus by inspecting
the $\vavg$ and $l_F$ shown in Fig.\ \ref{fig:props}(c,d),
we can study their roles in governing the conductivity.
Fermi-surface lengths and average Fermi-velocity
are similar in Ti$_3$C$_2$, Ti$_2$C, and Ti$_2$N.
Ti$_4$N$_3$ shows larger total length due to large number
of states crossing the Fermi-level,
but the average Fermi-velocities in these states are low.
Overall, our results indicate that by simply tuning the surface
composition the sheet conductivity cannot be changed by more than
about a factor of two.

Trends in electrical conductivity are often evaluated
from the density of states (DOS) at Fermi-level,
since $l_F \sim D_{\rm 2D}(E_F) \vavg$ (cf.\ Supplemental Material).
The DOS at Fermi-level are presented in \ref{fig:props}(e) and shows that 
the trends in the conductivity changes with composition
or between different materials cannot be extracted solely from the DOS
due to missing the effect of $\vavg$.

Experimentally measuring the sheet conductivity of MXene monolayers
is challenging, but the conductivity of MXene films
has been reported in several publications.
In the case of Ti$_3$C$_2$ films,
conductivities above $10^5$ S/m have been reported often
\cite{Halim14_CM,Ling14_PNAS,Shahzad16_Sci,Liu20_AEM}.
Recently a value as high as $1.1\cdot 10^6$ S/m was achieved
for a high crystalline quality Ti$_3$C$_2$ monolayer
\cite{Lipatov21_Mat}.
For other considered materials, the reports are more scarce,
but nevertheless fairly similar values can be found:
$5\cdot 10^5$ S/m for Ti$_2$N \cite{Soundiraraju17_ACSNano},
and $2.3\cdot 10^5$ S/m for Ti$_2$C \cite{Halim19_JPCM}.
It is worth noting, that
the conductivity may be governed by interflake contacts
and thus also sensitive to the presence of any intercalated species
\cite{Hart19_NComm}.

In order to estimate the film conductivity, we calculate
$\sigma = (\sigma_{\rm 2D}/\tau_0) \cdot \tau_0 / t$, where
$\tau_0$ is the relaxation time and 
$t$ is the interlayer separation taken from experiments:
10.3, 7.5,\cite{Naguib12_ACSNano} 14,\cite{Urbankowski16_Nanos} and 8.7 {\AA}\cite{Soundiraraju17_ACSNano}
for Ti$_3$C$_2$, Ti$_2$C, Ti$_4$N$_3$, and Ti$_2$N, respectively.
Note, that the interlayer separation depends on the preparation method,
as, e.g., intercalation of Li can increase the spacing by a few {\AA}.
Here we do not attempt to evaluate $\tau_0$ from
first-principles calculations, but instead
from the comparison to experimentally measured conductivities.
In Fig.\ \ref{fig:props}(b) we show the results for
$\tau_0 = 1$ fs, which yields $\sigma$ values comparable to
majority of the experimental reports.
The result of Ref.\ \onlinecite{Lipatov21_Mat} can be reproduced
by taking slightly larger $\tau\approx 2-3$ fs, depending on the surface composition.
The lower $\tau_0$ in earlier reports may then be ascribed
to larger number of defects caused by the strong etching,
although the interflake contacts may also contribute.

\subsection{Angle-resolved photoemission spectrum (ARPES)}

\begin{figure*}[!ht]
\begin{center}
  \includegraphics[width=16cm]{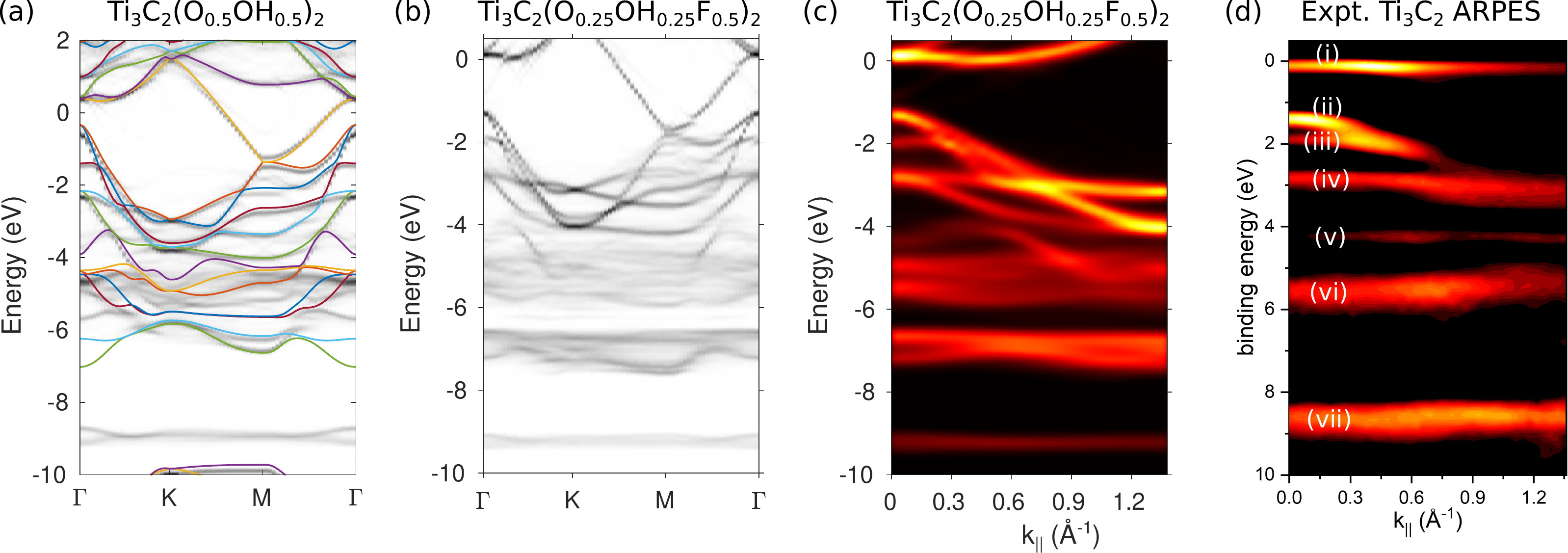}
\end{center}
\caption{\label{fig:arpes}
  (a) Calculated effective band structure
  of Ti$_3$C$_2$(O$_{0.5}$OH$_{0.5}$)$_2$ overlaid with
  the band structure from pseudohydrogenated system.
  (b) Effective band structure of
  Ti$_3$C$_2$(O$_{0.25}$OH$_{0.25}$F$_{0.5}$)$_2$.
  (c) The same as panel (b), but showing only the $\Gamma$-K line,
  broadened, and colormap chosen to match that in the experiments.
  (d) Experimentally measured ARPES spectrum of Ti$_3$C$_2$T$_x$,
  adapted from Ref.\ \onlinecite{Schultz19_CM}.
}
\end{figure*}

To verify that our calculations properly
capture the main features in the electronic structures of MXenes,
we compare them to the only available
experimental piece of information: the ARPES results
for Ti$_3$C$_2$ reported in Ref.\ \onlinecite{Schultz19_CM}.
The experimental data is reproduced in Fig.\ \ref{arpes}(d)
and shows: (i) flat band at the Fermi-level,
(ii) band with negative dispersion starting at -1.5 eV,
(iii) band with negative dispersion starting at -2 eV,
(iv) flat band at -3 eV,
(v) flat band segment at -4 eV,
(vi) broad flat band between -5 and -6 eV,
and (vii) broad flat band between -8 and -9 eV.
The composition was determined to be Ti$_3$C$_2$F$_{0.8}$O$_{0.8}$.

In Fig.\ \ref{fig:arpes}(a) we replot the effective band structure of
Ti$_3$C$_2$(O$_{0.5}$OH$_{0.5}$)$_2$
overlaid with the pseudohydrogenated band structure
from Fig.\ \ref{fig:ebs}(a), but in a wider energy range.
As discussed, the non-bonding Ti bands at and above 0 eV and the
Ti-C bonding bands between -4 and 0 eV are well reproduced
in the pseudohydrogenated systems.
The Ti-O bands between -7 and -4 eV are not reproduced very well,
and the O-H states at -9 eV are understandably poorly described
by pseudohydrogens.
Also some parts of the effective band structure are more strongly
smeared by the surface group disorder than others, whereas
the pseudohydrogen band structures contain no disorder.
These factors limit the use of the pseudohydrogenated structures
for reproducing ARPES results of deeper states.

Since the features close to Fermi-level
[features (i)--(iii) above] are properly described by pseudohydrogen
models, we can start by comparing them to
the band structures in Fig.\ \ref{fig:bs}.
Feature (i) indicates that the non-bonding band at $\Gamma$-point
must be occupied, and therefore O content must be less than about $0.25$.
The gap of about 1.5 eV between features (i) and (ii) would match
to the pure OH-terminated surface.
We then calculated effective band structures for few different mixtures
with small O content and found overall best agreement with
Ti$_3$C$_2$(O$_{0.25}$OH$_{0.25}$F$_{0.5}$)$_2$, as shown in
Fig.\ \ref{fig:arpes}(b)
and processed such that it is more comparable to the experimental spectrum
in Fig.\ \ref{fig:arpes}(c).
In this case, features (i)--(iv) are all reproduced very well.
The EBS also shows flat band away from the $\Gamma$-point
at about -4 eV, matching well with feature (v) in the experimental spectrum.
We also reproduce fairly well the broad band between -5 and -6 eV, feature (vi).
The lowest energy feature (vii) is not reproduced in our calculations.
The band at -7 eV in EBS originates from Ti-F bonds
and the band at below -9 eV from O-H bonds.
Since semilocal functionals, such as PBEsol one used here,
are known to underestimate band widths and experimentally measured
sample should contain significant amount of F, we ascribe
feature (vii) to Ti-F bonds,
in agreement with the discussion in Ref.\ \onlinecite{Schultz19_CM}.

\subsection{Conclusions}

In conclusion, we presented a study of the fermiology of
titanium carbide and nitride MXenes via density-functional theory
calculations.
We first calculated the effective band structures for surfaces
with mixtures of O and OH groups. 
The electronic structure near the Fermi-level was found to evolve
gradually with the composition, and we further showed they 
are very well reproduced by surfaces fully terminated with
pseudohydrogenated OH groups.
With the pseudohydrogenated surfaces we could readily
access Fermi-surface properties, such as Fermi-surface area
(or length in the case of 2D materials) and Fermi-velocity. 
We calculated the electrical conductivity within the constant
relaxation time approximation and by comparing to the experimentally
determined conductivities we obtained relaxation time of about 1 fs.
For the same relaxation time,
the conductivities for all studied MXenes were rather similar and
also fairly independent of the surface composition, changing
at most by a factor of two.
Finally, we compared our effective band structures to
those obtained from recent ARPES experiments \cite{Schultz19_CM}
for Ti$_3$C$_2$, where the best agreement was obtained using
Ti$_3$C$_2$(O$_{0.25}$OH$_{0.25}$F$_{0.5}$)$_2$ model.

The approaches demonstrated here can be used also for other MXenes,
including double MXenes \cite{Anasori15_ACSNano}
and even solid solutions in the M or X sublattice
if their positions and those of the surface functional
groups are uncorrelated.
With the pseudohydrogenation approach it should be possible to
calculate many other material properties of MXenes with
an effective mixed surface composition, such as optical properties
and electron-electron and electron-phonon scattering rates.
It will also allow investigating how instabilities
in electronic structure, leading to e.g. magnetism or charge density waves,
depend on the surface composition.


\section*{Acknowledgments}

We thank Prof.\ Norbert Koch and Dr.\ Thorsten Schultz for providing us
the original ARPES data, previously published in Ref.\ \onlinecite{Schultz19_CM}.
We are grateful to the Academy of Finland for support
under Academy Research Fellow funding No. 311058. We also thank
CSC--IT Center for Science Ltd. for generous grants of computer time.

\bibliography{../../../mxene,addref}

\end{document}


\title{Supplemental Material to ``Fermiology of two-dimensional titanium carbide and nitride MXenes''}

\author{Mohammad Bagheri$^1$}
\author{Rina Ibragimova$^2$}
\author{Hannu-Pekka Komsa$^{1,2}$}
\affiliation{
$^1$Microelectronics Research Unit, University of Oulu, Oulu, Finland \\
$^2$Department of Applied Physics, Aalto University, Aalto, Finland.
}

\date{\today}

\maketitle

\begin{figure*}[!h]
\begin{center}
  \includegraphics[width=12cm]{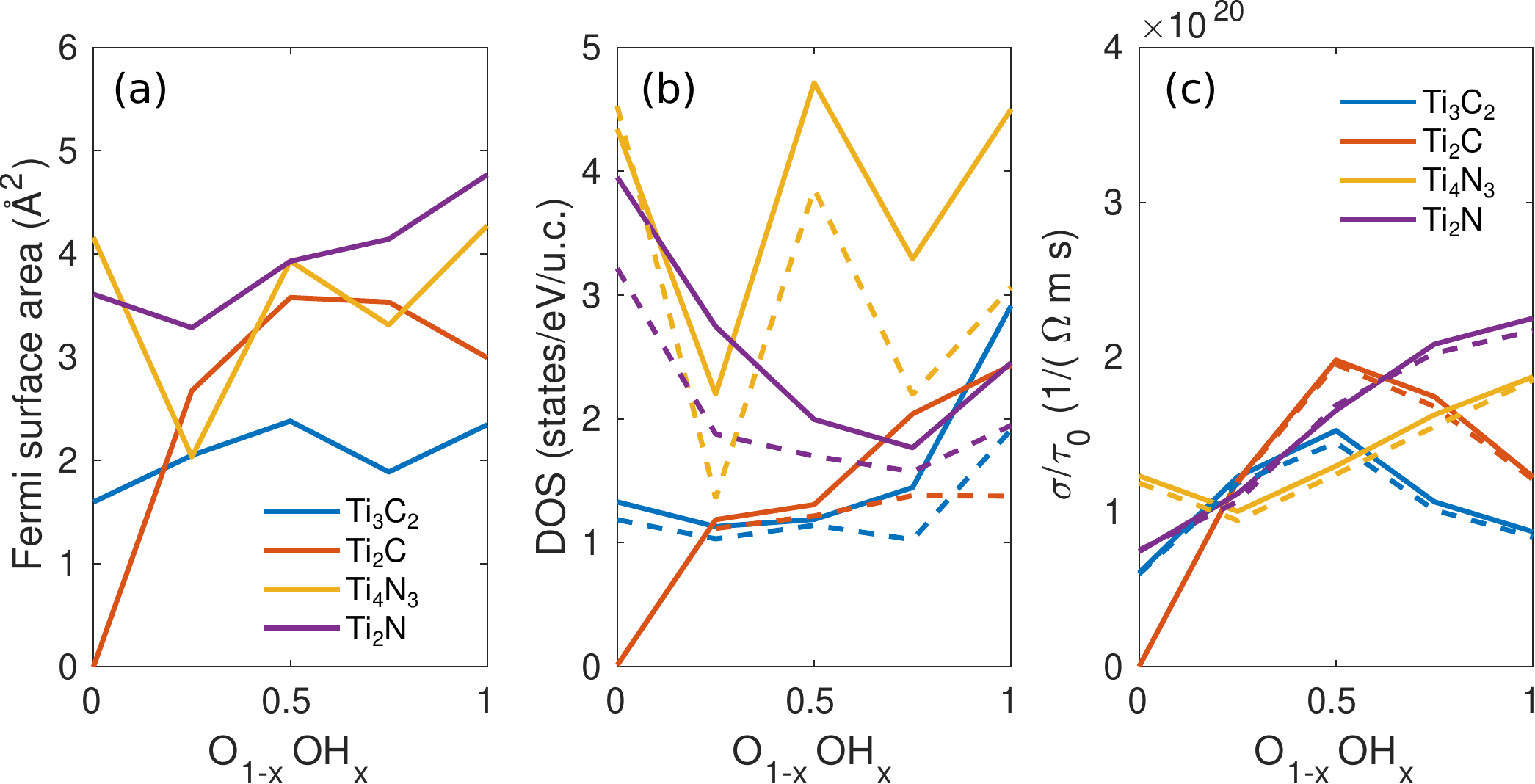}
\end{center}
\caption{\label{fig:props3D}
  Fermi-surface properties as obtained using the bulk (3D) equations:
  (a) Fermi-surface area, (b) density of states, and
  (c) electrical concductivity divided by relaxation time $\sigma/\tau_0$.
}
\end{figure*}

3D DOS at the Fermi-level is shown in Fig.\ \ref{fig:props3D}(b)
in the units of 1/eV/unitcell and including a factor of two for spin.
Within the free-electron model, it could also be calculated from
\begin{equation}
D(E_F) = \frac{2}{(2\pi)^3 \hbar}\cdot \frac{a_F}{\vavg} \cdot V_{\rm u.c.}
\end{equation}
where $a_F$ is the Fermi-surface area, $\vavg$ the average Fermi-velocity,
and $V_{\rm u.c.}$ the unit cell volume in real space.
This approximation is shown by dashed lines in Fig.\ \ref{fig:props3D}(b)
and shows that it holds fairly well when the Fermi-surface consists of
only one set of ellipsoids from a single band.
Note, that 3D DOS is equal to 2D DOS (given here for the free-electron model)
\begin{equation}
  D_{\rm 2D}(E_F) = \frac{2}{(2\pi)^2 \hbar}\cdot \frac{l_F}{\vavg} \cdot A_{\rm u.c.}
\end{equation}
where $l_F$ is the Fermi-surface length and $A_{\rm u.c.}$ the unit cell area.
This can be immediately verified by plugging in $l_F = a_F/b_z$ and $A_{\rm u.c.}=V_{\rm u.c.}/L_z$,
where $L_z$ the unit cell length in z-direction and $b_z=2\pi/L_z$ is
the reciprocal space basis vector in the out-of-plane direction.

The electrical conductivity is calculated from
\begin{subequations}
\begin{align}
  \sigma_{ij}
  &= \frac{e^2}{4\pi^3\hbar} \int_{E=E_F} \frac{v_i(k)v_j(k)}{|v(k)|}\tau(k) dk_F \label{eq:sigma1a}\\
  &\approx \frac{e^2}{4\pi^3\hbar} \tau_0 \int_{E=E_F} \frac{v_i(k)v_j(k)}{|v(k)|} dk_F \label{eq:sigma1b}\\
  &\approx \frac{e^2}{4\pi^3\hbar} \tau_0 \frac{\vavg}{2} \cdot a_F \label{eq:sigma1c}
\end{align}
\end{subequations}
We have assumed $\tau(k)$ is constant $\tau_0$
and $\vavg a_f = \int |v| = \int v_x^2/|v| + \int v_y^2/|v| \approx 2 \int v_x^2/|v|$.

The sheet conductivity is written similarly:
\begin{subequations}
\begin{align}
  \sigma_{\rm 2D}
  &= \frac{e^2}{2\pi^2\hbar} \tau_0 \int_{E=E_F} \frac{v_i(k)v_j(k)}{|v(k)|} dk_F \label{eq:sigma2a} \\
  &= \frac{e^2}{2\pi^2\hbar} \tau_0 \frac{\vavg}{2} \cdot l_F \label{eq:sigma2b} \\
  &= \sigma \cdot \frac{2\pi}{b_z} = \sigma \cdot L_z \label{eq:sigma2c}
\end{align}
\end{subequations}

We first use Boltztrap2 to calculate $\sigma$ and then use Eq.\ \ref{eq:sigma2c}
to get $\sigma_{\rm 2D}$,
where $L_z$ is the length of the lattice constant $c$ of the computational cell.
In order to evaluate $\sigma$ of MXene films, we again use  Eq.\ \ref{eq:sigma2c}, but this
time $L_z$ is the layer separation.

The ``3D quantities'' obtained for the monolayer are
shown in Fig.\ \ref{fig:props3D}, but depend on the size of the vacuum region.
The cell volume and lattice constant $c$, which are required to translate
the 2D and 3D conductivities, are listed in Table \ref{tab:lattice}.

\begin{table*}[!h]
\begin{center}
\caption{\label{tab:lattice}
  Unit cell volumes $V$ (in {\AA}$^3$) and the lattice constants in z-direction $c$
  (in {\AA}) for all considered systems.
}
\begin{tabular}{lcccccccccc}
  & \multicolumn{2}{c}{O} & \multicolumn{2}{c}{O$_{0.75}$OH$_{0.25}$} 
  & \multicolumn{2}{c}{O$_{0.5}$OH$_{0.5}$} & \multicolumn{2}{c}{O$_{0.25}$OH$_{0.75}$} & \multicolumn{2}{c}{OH} \\
system          & V  & c     & V        & c    & V       & c      & V      & c      & V      & c \\
\hline

Ti$_3$C$_2$ & 232.41  & 29.53 & 242.71  & 30.97 & 242.71  & 30.70 & 242.71  & 30.30 & 242.71  & 29.98 \\
Ti$_2$C     & 154.70  & 19.75 & 154.70  & 19.96 & 154.70  & 19.78 & 154.70  & 19.54 & 154.70  & 19.34 \\
Ti$_4$N$_3$ & 258.45  & 33.58 & 258.45  & 33.61 & 258.45  & 33.46 & 258.45  & 33.33 & 258.45  & 33.20 \\
Ti$_2$N     & 154.70  & 20.20 & 154.70  & 20.21 & 154.70  & 19.99 & 154.70  & 19.71 & 154.70  & 19.52 \\
\end{tabular}
\end{center}
\end{table*}

